\documentclass[12pt]{article}
\usepackage{amsmath}
\usepackage{amssymb}
\usepackage[english]{babel}
\usepackage{latexsym}
\usepackage{times}
\usepackage{mathptm}
\begin{document}
\small
\normalsize
\newcounter{saveeqn}
\newcommand{\alpheqn}{\setcounter{saveeqn}{\value{equation}}%
\setcounter{equation}{0}%
\renewcommand{\theequation}{\mbox{\arabic{saveeqn}-\alph{equation}}}}
\newcommand{\reseteqn}{\setcounter{equation}{\value{saveeqn}}%
\renewcommand{\theequation}{\arabic{equation}}}

\protect\newtheorem{principle}{Principle} 
\protect\newtheorem{theo}[principle]{Theorem}
\protect\newtheorem{prop}[principle]{Proposition}
\protect\newtheorem{lem}[principle]{Lemma}
\protect\newtheorem{co}[principle]{Corollary}
\protect\newtheorem{de}[principle]{Definition}
\newtheorem{ex}[principle]{Example}
\newtheorem{rema}[principle]{Remark}
\newtheorem{state}[principle]{Statement}{\bf}{\rm}
\newtheorem{acknowledgements}[principle]{Acknowledgements}{\bf}{\rm}
\vspace*{2.6cm}
\noindent {\Large \textsf{A uniqueness theorem for entanglement measures} } \\ \\ \\ \\
{\large {Oliver Rudolph}}  \\ \vskip.0001in
\noindent {\normalsize
Physics Division, Starlab nv/sa, Engelandstraat 555, B-1180
Brussels, Belgium. \\}
{\normalsize E-mail: rudolph@starlab.net}
\\
\\ \\ \\
\normalsize
\noindent \emph{Abstract} We obtain a mathematically simple characterization of all
functionals coinciding with the von Neumann reduced entropy on
pure states based on the Khinchin-Faddeev axiomatization of Shannon entropy
and give a physical interpretation of the axioms in
terms of entanglement.
\\ \\
\section{Introduction} The characterization and classification of
entanglement in quantum mechanics is one of the
cornerstones of the emerging field of quantum information theory.
This note is devoted to the study of entanglement measures.
Entanglement measures are positive real-valued
functionals defined on the joint state space
of two or more quantum systems (subject to further requirements).
A number of entanglement measures have been discussed in the
literature, such as the von Neumann reduced entropy, the relative
entropy of entanglement \cite{PlenioV98},
the so-called entanglement of distillation
and the entanglement of formation \cite{BennettVSW97}.
Several authors proposed physically motivated postulates
to characterize entanglement
measures (see, e.g., \cite{PlenioV98,VedralPRK97,Horodecki99}). These postulates (although they vary from author to
author in the details) have in common that they are
based on the concepts of the operational formulation of quantum mechanics
\cite{Kraus83}.
Mathematically, the list of postulates
for entanglement measures
serves as the definition of the notion of entanglement measure.
Many authors agree that
the only physically reasonable entanglement measure on pure states
is given by the von Neumann reduced entropy.
Indeed, it is a known fact that there are important
entanglement measures which do coincide with the von Neumann reduced
entropy on pure states, for instance the relative
entropy of entanglement \cite{PlenioV98}. Accordingly, it is
generally seen as a desirable property of the axiomatic
characterization of entanglement measures that it allows only for
entanglement measures which coincide with the von Neumann reduced
entropy on pure states. This point of view is supported by an argument by
Popescu and Rohrlich \cite{PopescuR97}. However,
their uniqueness theorem was not rigorously proved under unified
assumptions and definitions.
A mathematically exact
statement and proof of the \emph{operational} uniqueness theorem
has recently been given in Ref.~\cite{DonaldHR01}.

In this paper we present an alternative list of
physically reasonable and mathematically simple
postulates for entanglement measures such that
all entanglement measures satisfying these postulates coincide
with the von Neumann reduced entropy on pure
states. Mathematically our postulates are just
an adaptation of the Khinchin-Faddeev characterization of Shannon
entropy \cite{Khinchin57} and all proofs in this paper are elementary.
The main purpose of this note is to not to communicate new mathematical
methods but rather that physically
the Khinchin-Faddeev postulates
admit a natural interpretation in terms of entanglement rather than
information or lack of information.
What is also interesting about the
present result is firstly that our axiomatization can be formulated
without resorting to the mathematical apparatus and the physical concepts
of the theory of
local quantum operations and secondly
that the technical asymptotic
requirements which are so central to the Popescu-Rohrlich
argument (see \cite{PopescuR97,DonaldHR01}) are not needed as well.
It turns out that they can be replaced
basically by a single intuitive and comparably
weak requirement which fixes
the value of entanglement measures on pure states.

\section{Preliminaries}
In this section we collect some basic definitions and results
which are used in the course of this paper.

In the present paper we restrict ourselves mainly to the situation of
a composite quantum system consisting of two subsystems with Hilbert space
${\mathcal{H}}_1 \otimes {\mathcal{H}}_2$ where ${\mathcal{H}}_1$ and
${\mathcal{H}}_2$ denote the Hilbert spaces of the subsystems.
The states of the system are identified with the density operators
on ${\mathcal{H}}_1 \otimes {\mathcal{H}}_2$. A density operator is a
positive trace class operator with trace one.
\begin{de}
Let ${\cal H}_1$ and ${\cal H}_2$ be two Hilbert spaces of
arbitrary dimension. A density operator $\varrho$ on the tensor product
${\cal H}_1
\otimes {\cal H}_2$ is called \emph{separable} or
\emph{disentangled} if there exist a
family $\left\{ \omega_{i} \right\}$ of positive real numbers, a family
$\left\{ \rho^{(1)}_i \right\}$ of density operators on
${\cal H}_1$ and a family $\left\{ \rho^{(2)}_i \right\}$ of
density operators
on ${\cal H}_2$ such that
\begin{equation} \label{e1}
\varrho = \sum_{i} \omega_{i} \rho^{(1)}_i \otimes \rho^{(2)}_i,
\end{equation}
where the sum converges in trace class norm.
\end{de}
The set of states is a convex set and its extreme points,
which are also called \emph{pure states}, are the
projection operators. Every pure state obviously corresponds to a
unit vector $\psi$ in ${\mathcal{H}}_1 \otimes {\mathcal{H}}_2$.
We denote the projection operator onto the subspace spanned by the
unit vector $\psi$ by $P_\psi$.

The Schmidt decomposition (see also \cite{EkertK90})
is of central importance in the
characterization and quantification of entanglement associated
with pure states.
\begin{lem}
Let ${\mathcal{H}}_1$ and ${\mathcal{H}}_2$ be Hilbert spaces of
arbitrary dimension and let $\psi \in {\mathcal{H}}_1 \otimes
{\mathcal{H}}_2$.
Then there exist a family of non-negative real numbers $\{
p_i \}_i$ and orthonormal bases $\{ a_i \}_i$ and $\{ b_i \}_i$ of
${\mathcal{H}}_1$ and ${\mathcal{H}}_2$ respectively such that
\[ \psi = \sum_i \sqrt{p_i} a_i \otimes b_i. \] \label{Sch}
\end{lem}
The family of positive numbers $\{ p_i \}_i$ is called the family
of \emph{Schmidt coefficients} of $\psi$.
For pure states the family of Schmidt coefficients of a state
completely characterizes the amount of entanglement of that state.
A pure state $\psi$ is separable if and only if $\psi = a \otimes
b$ for some $a \in {\mathcal{H}}_1$ and $b \in {\mathcal{H}}_2$.
With every vector $\psi$
in ${\mathcal{H}}_1 \otimes {\mathcal{H}}_2$ we associate
a closed subspace $M(\psi)$ of ${\mathcal{H}}_1 \otimes {\mathcal{H}}_2$:
let $\psi = \sum_i \sqrt{p_i} a_i \otimes b_i$ be the Schmidt decomposition
of $\psi$ as in Lemma \ref{Sch}, then $M(\psi)$ is defined as
the subspace of ${\mathcal{H}}_1 \otimes {\mathcal{H}}_2$
spanned by all simple product states $a_i \otimes b_i$
with nonzero Schmidt coefficient in the Schmidt decomposition of $\psi$.
We call $M(\psi)$ the \emph{Schmidt subspace} associated with $\psi$.
The dimension of $M(\psi)$ is called the \emph{Schmidt rank} of
$\psi$. Moreover, we call
two states $\psi$ and $\phi$ \emph{Schmidt orthogonal} if
$M(\psi)$ and $M(\phi)$ are orthogonal.

The \emph{von Neumann reduced entropy}
for density operators
$\sigma$ on a tensor product Hilbert space ${\mathcal{H}}_1 \otimes
{\mathcal{H}}_2$ is defined as
\begin{equation}
S_{\mathrm{vN}}(\sigma) := -
{\mathrm{Tr}}_{{\mathcal{H}}_1}({\mathrm{Tr}}_{{\mathcal{H}}_2}
\sigma \ln ({\mathrm{Tr}}_{{\mathcal{H}}_2}
\sigma)),
\end{equation} where
${\mathrm{Tr}}_{{\mathcal{H}}_1}$ and ${\mathrm{Tr}}_{{\mathcal{H}}_2}$
denote the partial traces over
${\mathcal{H}}_1$ and ${\mathcal{H}}_2$ respectively. In the case of
pure states $\sigma = P_\psi$, it can be shown that
$- {\mathrm{Tr}}_{{\mathcal{H}}_1}({\mathrm{Tr}}_{{\mathcal{H}}_2}
P_\psi
\ln ({\mathrm{Tr}}_{{\mathcal{H}}_2} P_\psi)) = -
{\mathrm{Tr}}_{{\mathcal{H}}_2}({\mathrm{Tr}}_{{\mathcal{H}}_1} P_\psi
\ln ({\mathrm{Tr}}_{{\mathcal{H}}_1} P_\psi)) = - \sum_i p_i \ln p_i$
where $\{ p_i \}_i$ denotes the family of Schmidt coefficients of $\psi$.
However, for a general mixed state $\sigma$ we have
${\mathrm{Tr}}_{{\mathcal{H}}_1}({\mathrm{Tr}}_{{\mathcal{H}}_2} \sigma
\ln ({\mathrm{Tr}}_{{\mathcal{H}}_2} \sigma)) \linebreak[3] \neq
{\mathrm{Tr}}_{{\mathcal{H}}_2}({\mathrm{Tr}}_{{\mathcal{H}}_1} \sigma
\ln ({\mathrm{Tr}}_{{\mathcal{H}}_1} \sigma))$.

The Khinchin-Faddeev axiomatization of Shannon entropy can be
formulated as follows (taken from \cite{OhyaP93})
\begin{lem}
Let $\Pi$ denote the set of all probability distributions $(p_1,
\cdots, p_n)$ ($p_i \geq 0, \sum_i p_i =1$).
Let $S: \Pi \to {\mathbb{R}}$ be
a function satisfying
\begin{itemize}
\item Continuity: $p \mapsto S(p, 1-p)$ is continuous on $[0,1]$.
\item Normalization: $S(1/2,1/2) = \log 2$.
\item Symmetry: $S(p_{\kappa(1)}, \cdots, p_{\kappa(n)}) = S(p_1, \cdots,
p_n)$ for all permutations $\kappa$ of $\{ 1, \cdots, n \}$.
\item Recursion: For every $0 \leq \eta \leq 1$ we have $S(p_1, \cdots,
p_{n-1}, \eta p_n, (1- \eta) p_n) = S(p_1, \cdots, p_n) + p_n S(\eta, 1 -
\eta).$
\end{itemize} Then $S$ is equal to the Shannon entropy, i.e.,
$S(p_1, \cdots, p_n) = - \sum_{i=1}^n p_i \log p_i$.
\end{lem}
\section{Khinchin-Faddeev-type postulates for entanglement measures}
\subsection{Entanglement measures on pure states}
Let ${{\mathcal{H}}_1}$ and ${{\mathcal{H}}_2}$ be Hilbert spaces.
For the
moment we restrict ourselves to the problem of characterizing
entanglement measures defined on pure states, i.e., functionals
$E : {{\mathcal{H}}_1} \otimes {{\mathcal{H}}_2} \to {\mathbb{R}}^+$.
The discussion in this section starts from the question: what
are the minimal conditions we want to impose on
a mathematically satisfactory measure of entanglement?
It is reasonable to require that $E$ is defined and continuous on
the tensor product of the Hilbert spaces of any given two systems.
Moreover, we have argued above that the sequence of Schmidt
coefficients fully characterizes the entanglement of a pure state.
Therefore we expect $E(\psi)$ to depend only on the Schmidt coefficients
of $\psi$. Equivalently, we require that
for any given system ${{\mathcal{H}}_1} \otimes {{\mathcal{H}}_2}$
the entanglement measure $E : {{\mathcal{H}}_1} \otimes {{\mathcal{H}}_2}
\to {\mathbb{R}}^+$ is invariant under unitary operations of the
form $U \otimes V$, where $U$ and $V$ are unitaries on
${{\mathcal{H}}_1}$ and
on ${{\mathcal{H}}_2}$ respectively. We also require that $E$ is
invariant under embeddings into larger Hilbert spaces.
\begin{itemize}
\item[(P1)] An entanglement measure is a positive real-valued functional
$E$ which for any given two systems is well-defined
on the tensor product of the Hilbert spaces of the
two systems. For any given two systems corresponding to Hilbert
spaces ${{\mathcal{H}}_1}$ and ${{\mathcal{H}}_2}$ the function
${{\mathcal{H}}_1} \otimes {{\mathcal{H}}_2} \ni \psi \mapsto E(\psi)$ is
continuous in the norm topology.
\item[(P2)] For any given two systems with Hilbert
spaces ${{\mathcal{H}}_1}$ and ${{\mathcal{H}}_2}$ the function
$E: {{\mathcal{H}}_1} \otimes {{\mathcal{H}}_2} \to {\mathbb{R}}^+$ satisfies
\[ E(U \otimes V \psi) = E(\psi) \] for all $\psi \in {{\mathcal{H}}_1}
\otimes {{\mathcal{H}}_2}$ and all unitaries $U,V$ acting on
${{\mathcal{H}}_1}$ and ${{\mathcal{H}}_2}$ respectively.
\item[(P3)] whenever $\psi \in {{\mathcal{H}}_1} \otimes {{\mathcal{H}}_2}
\subset {\mathsf{H}}_1 \otimes {\mathsf{H}}_2$ with embeddings
${\mathcal{H}}_1 \hookrightarrow {\mathsf{H}}_1$ and
${{\mathcal{H}}_2} \hookrightarrow {\mathsf{H}}_2$
of ${\mathcal{H}}_1$ and ${{\mathcal{H}}_2}$ into larger Hilbert spaces
${\mathsf{H}}_1$ and ${\mathsf{H}}_2$ respectively, then
$E \vert_{{{\mathcal{H}}_1} \otimes {{\mathcal{H}}_2}}(\psi) =
E \vert_{{\mathsf{H}}_1 \otimes {\mathsf{H}}_2}(\psi)$.
\end{itemize}
We exclude the identically vanishing functional $E \equiv
0$.
It is an immediate consequence of (P2) and (P3), that for every pure state
$\psi$ the value $E(\psi)$ does only depend on the non-zero
Schmidt coefficients of $\psi$. We will therefore also write
$E(\lambda_1, \cdots, \lambda_n)$ for $E(\psi)$ where $\{ \lambda_1, \cdots,
\lambda_n \}$ denotes the family of (non-vanishing) Schmidt coefficients of
$\psi$.

Hilbert spaces are linear spaces, and therefore, trivially,
all normalized linear
combinations of pure states are pure states themselves. Consider
for instance $\phi = \sum_m \sqrt{p_m} \psi_m$ where $\{p_m \}$
is a probability distribution and where
$\psi_m \in {{\mathcal{H}}_1} \otimes {{\mathcal{H}}_2}$.
Naively one might hope that the entanglement $E(\phi)$ of $\phi$
is a weighted sum of the entanglement of the $\psi_m$ plus the
entanglement associated with the superposition $E(p_1, \cdots,
p_m)$. It is easy to
see that this is not true in general, and that on the contrary
superpositions of maximally entangled states can even be unentangled.
Mathematically this can be traced back to the fact that the
Schmidt spaces of the superimposed states are not orthogonal.
Generally, superposing a family $\psi_1, \cdots,
\psi_m$ of pure states whose Schmidt spaces are not mutually orthogonal
may increase as well as decrease entanglement. However,
in the case of superpositions of mutually Schmidt orthogonal pure states
no term in the Schmidt decomposition of one state can cancel
terms in the Schmidt decomposition of another state.
Therefore
for the special case of a superposition $\phi = \sum_m \sqrt{p_m} \psi_m$
of a family $\{ \psi_1, \cdots, \psi_m \}$
of mutually Schmidt orthogonal pure states (where
$\{ p_1, \cdots, p_m \}$ is a probability distribution), what we
expect physically is that the
entanglement of the superposed state equals the averaged
entanglement of the $\{ \psi_i \}$ plus the entanglement
$E(p_1, \cdots, p_n)$ associated with the superposition. More formally we
require
\begin{itemize}
\item[(P4)] Let $\{ \psi_1, \cdots, \psi_m \}$ be a family of
mutually Schmidt orthogonal pure states and $\{ \lambda_1, \cdots,
\lambda_m \}$ be a distribution of probability amplitudes, i.e., a
sequence of complex numbers with \\
$\sum_{i=1}^m \vert \lambda_i \vert^2
=1$, then \[ E(\lambda_1 \psi_1 + \cdots + \lambda_m \psi_m) =
E(\vert \lambda_1
\vert^2, \cdots, \vert \lambda_m \vert^2 ) +
\sum_{i=1}^m \vert \lambda_i \vert^2 E(\psi_i). \]
\end{itemize}
We note the following
\begin{lem} \label{sep1}
Let $E$ be an entanglement measure on pure states satisfying (P1),
(P2), (P3) and (P4). Then $E(\psi) = 0$ for all separable pure states.
\end{lem}
\emph{Proof}: Let $\psi_1, \psi_2$ be two separable orthogonal
pure states.
Then by (P4)
$E(1 \psi_1 + 0 \psi_2) = E(1,0) = 1 E(\psi_1) + 0 E(\psi_2) +
E(1,0)$. Thus $E(\psi_1) = 0$. $\Box$

\begin{lem}
The von Neumann reduced entropy $S_{\mathrm{vN}}$ satisfies (P1),
(P2), (P3) and (P4). \end{lem}
\emph{Proof}: Straightforward. $\Box$ \\

We show that the requirements (P1) - (P4) already fix
the von Neumann reduced entropy up to a multiplicative constant.
\begin{prop}
Let $E$ be an entanglement measure on pure states satisfying the
postulates (P1), (P2), (P3) and (P4). Then there exist a positive real
constant $c$ such that $E = c S_{\mathrm{vN}}$.
\end{prop}
\emph{Proof}: Let $\psi_1, \cdots, \psi_{n+1}$ be a collection of
mutually orthogonal separable pure states and $\{ \lambda_1,
\cdots, \lambda_{n} \}$ be a distribution of complex probability amplitudes
and $\eta \in [0,1]$.
Then \begin{eqnarray*} & &
E(\lambda_1 \psi_1 + \cdots + \lambda_{n-1} \psi_{n-1} +
\sqrt{\eta}
\lambda_n \psi_n + \sqrt{1 - \eta} \lambda_n \psi_{n+1}) = \\ & &  = \vert
\lambda_n \vert^2 E(\sqrt{\eta} \psi_n + \sqrt{1- \eta} \psi_{n+1}) +
(1 - \vert \lambda_n \vert^2) E\left( \frac{1}{1 - \vert \lambda_n \vert^2}
(\lambda_1 \psi_1 + \cdots + \lambda_{n-1} \psi_{n-1}) \right) +
\\ & & \, \, \, \, \, \, \, + E(\vert \lambda_n \vert^2, 1 - \vert \lambda_n \vert^2).
\end{eqnarray*}
Moreover, \begin{eqnarray*} E(\lambda_1 \psi_1 + \cdots + \lambda_n \psi_n)
& =
& \vert \lambda_n \vert^2 E(\psi_n) + (1 - \vert \lambda_n \vert^2)
E \left( \frac{1}{1- \vert \lambda_n \vert^2} (\lambda_1 \psi_1 +
\cdots + \lambda_{n-1} \psi_{n-1}) \right) + \\ & & +
E(\vert \lambda_n \vert^2,
1 - \vert \lambda_n \vert^2). \end{eqnarray*} Thus
\[E(\vert \lambda_1 \vert^2, \cdots, \vert \lambda_{n-1} \vert^2,
\eta \vert \lambda_n \vert^2, (1 - \eta) \vert \lambda_n \vert^2) =
E(\vert \lambda_1 \vert^2, \cdots, \vert \lambda_n \vert^2) + \vert \lambda_n
\vert^2 E(\eta, 1- \eta).\] Therefore $E$ considered as a function of
the Schmidt coefficients satisfies all conditions of the
Khinchin-Faddeev characterization of Shannon's entropy. Therefore
$E(p_1, \cdots, p_n) = - c \sum_{i=1}^n p_i \ln p_i$ for some positive real
constant $c.$ $\Box$
\subsection{Entanglement measures for mixed states}
An entanglement measure on mixed states is a functional defined on
the state space of any given two quantum systems. If the Hilbert
spaces of the two systems are ${{\mathcal{H}}_1}$ and ${{\mathcal{H}}_2}$,
then the state space is the set of density operators on ${{\mathcal{H}}_1}
\otimes {{\mathcal{H}}_2}$, denoted by ${\mathcal{D}}({{\mathcal{H}}_1} \otimes
{{\mathcal{H}}_2}).$ An entanglement measure is then a functional satisfying the
obvious generalizations of (P1)-(P3)
\begin{itemize}
\item[(M1)] An entanglement measure is a positive real-valued functional
$E$ which for any given two systems is well-defined
on the set of density operators on the
tensor product of the Hilbert spaces of the
two systems. For any given two systems corresponding to Hilbert
spaces ${{\mathcal{H}}_1}$ and ${{\mathcal{H}}_2}$ the function
${\mathcal{D}}({{\mathcal{H}}_1} \otimes {{\mathcal{H}}_2}) \ni \rho \mapsto E(\rho)$
is continuous with respect to the trace class norm.
\item[(M2)] For any given two systems with Hilbert
spaces ${{\mathcal{H}}_1}$ and ${{\mathcal{H}}_2}$ the function
$E: {\mathcal{D}}({{\mathcal{H}}_1} \otimes {{\mathcal{H}}_2}) \to {\mathbb{R}}^+$
satisfies
\[ E(U \otimes V \rho U^\dagger \otimes V^\dagger) = E(\rho) \]
for all $\rho \in {\mathcal{D}}({{\mathcal{H}}_1}
\otimes {{\mathcal{H}}_2})$ and all unitaries $U,V$ acting on
${{\mathcal{H}}_1}$ and ${{\mathcal{H}}_2}$ respectively.
\item[(M3)] whenever $\rho \in {\mathcal{D}}({{\mathcal{H}}_1} \otimes
{{\mathcal{H}}_2}) \subset {\mathcal{D}}({\mathsf{H}}_1 \otimes
{\mathsf{H}}_2)$ with embeddings
${{\mathcal{H}}_1} \hookrightarrow {\mathsf{H}}_1$ and
${{\mathcal{H}}_2} \hookrightarrow {\mathsf{H}}_2$
of ${{\mathcal{H}}_1}$ and ${{\mathcal{H}}_2}$ into larger Hilbert spaces
${\mathsf{H}}_1$ and ${\mathsf{H}}_2$ respectively, then
$E \vert_{{{\mathcal{H}}_1} \otimes {{\mathcal{H}}_2}}(\rho) =
E \vert_{{\mathsf{H}}_1 \otimes {\mathsf{H}}_2}(\rho)$.
\end{itemize}
Moreover we require that (P4) is satisfied without change and that
mixing of states does not increase entanglement.
\begin{itemize}
\item[(M4)] Let $\{ \psi_1, \cdots, \psi_m \}$ be a family of
mutually Schmidt orthogonal pure states and $\{ \lambda_1, \cdots,
\lambda_m \}$ be a distribution of probability amplitudes,
then \[ E(P_\psi) =
E(\vert \lambda_1
\vert^2, \cdots, \vert \lambda_m \vert^2 ) +
\sum_{i=1}^m \vert \lambda_i \vert^2 E(P_{\psi_i}) \] where
$\psi \equiv \lambda_1 \psi_1 + \cdots + \lambda_m \psi_m$ and
where $P_\psi$ and $P_{\psi_i}$
denote the projection operators onto the subspace
spanned by $\psi$ and $\psi_i$ respectively.
\item[(M5)] Mixing of states does not increase entanglement, i.e.,
$E$ is convex \[ E(\eta \sigma + (1 - \eta) \tau) \leq
\eta E(\sigma) + (1 - \eta) E(\tau) \] for all $0 \leq
\eta \leq 1$ and all $\sigma, \tau \in {\mathcal{D}}({{\mathcal{H}}_1}
\otimes {{\mathcal{H}}_2}).$
\end{itemize}
\begin{lem}
Let $E$ be an entanglement measure on mixed states satisfying
(M1), (M2), (M3), (M4) and (M5). Then $E(\rho) = 0$ for all
separable states $\rho$. \label{sep2}
\end{lem}
\emph{Proof}: By Lemma \ref{sep1} $E$ vanishes for all separable
pure states. Every separable state $\rho$ is a statistical mixture
$\rho = \sum_{i=1}^n p_i P_{\psi_i}$ where $\{ \psi_i \}_{i=1}^n$ is a family
of separable pure states and where $(p_1, \cdots, p_n)$ is a
probability distribution (with $n$ possibly infinite).
Thus, by (M5) and (M1) \[ E(\rho) \leq
\sum_{i=1}^n p_i E(P_{\psi_i}) = 0. \] Hence $E(\rho) = 0$ for all
separable states $\rho$. $\Box$
\begin{ex} Let ${\mathsf{H}}$ and
${\mathsf{K}}$ be finite dimensional Hilbert spaces.
The greatest cross norm on the space of trace class operators
${\mathtt{T}}({\mathsf{H}} \otimes {\mathsf{K}})$ on ${\mathsf{H}} \otimes
{\mathsf{K}}$ is defined by
\begin{equation}
\Vert \sigma \Vert_\gamma := \inf \left\{ \sum_{i=1}^n
\left\Vert x_i \right\Vert_1 \, \left\Vert
{y}_i \right\Vert_1 \, \left\vert \, \sigma = \sum_{i=1}^n x_i
\otimes {y}_i \right. \right\}, \end{equation} where $\sigma \in
{\tt T}({\sf H} \otimes {\sf K})$, where the infimum runs over
all finite decompositions of $\sigma$ into elementary tensors and where
$\Vert \cdot \Vert_1$ denotes the trace class norm.
For projection operators $P_\psi$ the value of $\Vert \cdot
\Vert_\gamma$ has been computed in \cite{Rudolph00b}: $\Vert P_\psi
\Vert_\gamma = \left( \sum_i \sqrt{p_i} \right)^2$ where $\{ p_i
\}$ denotes the family of Schmidt coefficients of the unit vector
$\psi$. The entanglement measure introduced in \cite{Rudolph00b}
\[ E(\sigma) \equiv \Vert \sigma \Vert_\gamma \ln \Vert \sigma
\Vert_\gamma \] does not satisfy (P4).
\end{ex}
\section{Discussion} In this work we gave a mathematical
characterization of all functionals defined on the state space of
composite quantum systems which coincide with von Neumann reduced
entropy on pure states: a functional on pure states coincides with the von
Neumann reduced entropy if and only if it satisfies the conditions
(P1)-(P4). Mathematically the axioms (P1)-(P4) are just a
version of the Khinchin-Faddeev characterization of Shannon
entropy but physically we have seen that they admit an interpretation
in terms of entanglement. \\

\end{document}